\documentclass[prl,twocolumn,showpacs,floatfix,preprintnumbers,amsmath,amssymb]{revtex4}
\usepackage{graphicx}
\usepackage{dcolumn}
\usepackage{bm}
\pdfoutput=1

\begin{document}

\newcommand{\bec}{\begin{center}}
\newcommand{\ec}{\end{center}}
\newcommand{\be}{\begin{equation}}
\newcommand{\ee}{\end{equation}}
\newcommand{\beqn}{\begin{eqnarray}}
\newcommand{\eeqn}{\end{eqnarray}}
\newcommand{\bet}{\begin{table}}
\newcommand{\ent}{\end{table}}
\newcommand{\bib}{\bibitem}

\setlength{\textwidth}{18cm} 
\setlength{\textheight}{25.5cm} 

\baselineskip 4mm


\title{
Fe impurity induced ion-nanopatterning: atomistic 
simulations using a new force field for FeSi
}

\author{P. S\"ule} 
\affiliation
  {Research Institute for Technical Physics and Material Science,\\
Konkoly Thege u. 29-33, Budapest, Hungary,sule@mfa.kfki.hu,www.mfa.kfki.hu/$\sim$sule,\\
}

\date{\today}


\begin{abstract}
 The ion-bombardment induced nanopatterning of Si(001) has been
simulated by atomistic simulations with and without Fe impurity.
The surface contamination has been simulated by using a new force field
developed for FeSi. This is a fitted bond order potential (BOP) given
for Si and for Fe by Albe et al. This BOP formula has been
optimized simultaneously for FeSi, Si and for Fe.
Using this new force field we are able to follow
the ion-beam assisted deposition for Si in the presence of Fe contamination
in the surface region.
As an overall result, we get an unexpectedly rich
variety of nanopatterns formed by the reorganization of the crater rims 
of the individual ion impacts.
The previously thought simple atomistically roughened surfaces
show unprecedented landscapes and topography with nanoscale features.
The characteristic size of the units of the pattern is in the range of
a few nms.
Typical of the occurred pattern is the network of interconnected
elongated adatom islands.
We also see the self-organization of this pattern upon ion-bombardments.
At $50^{\circ}$ impact angle we get a nanoporous surface (sponge-like) both
for Fe-contaminated and Fe-free simulations.
At $70^{\circ}$ of impact (grazing angle of incidence)
the pattern resembles to that of elongated atomic chains (adatom islands)
along the ion impact direction.
This latter pattern could be understood as a prepattern state towards
rippling.
At lower angles ($30^{\circ}$) nanoholes rule the landsdcape.
The obtained pattern corresponds to low fluence experiments which are
used to consider as simple roughening without showing any sing of patterning.

\end{abstract}
\maketitle

\section{Introduction}

 Self-assembly, that is the spontaneous formation of regular structures, has become a critical implement in the toolbox of nanotechnologists.
 Self-organising functional nano-systems and nano-devices are the ultimate aim of bottom-up nano-fabrication \cite{bottom}.
The spontaneous organisation can be exploited for nanopatterning (NP) a variety of materials
using a variety of nanotechnological tools \cite{bottom}.
Affordable surfaces with well-controlled nanostructures over a large area open new applications not only in electronics but also in the physical world through their unique properties originating from their nanoscale geometry.
The direct nanofabrication of the surface (top-down approach) using coupled interference lithography with deep reactive ion etching also become an efficient tool recently for nanofabrication,
although, the required number of technological steps could be still too large.
\cite{nanopost,nanolithog}.

 So far much of the work related to self-assembling nanostructures (bottom-up) has been nothing more than demonstrations in university laboratories.
Therefore, the builiding up of regular nanopatterns using the
spontaneous formation of surface features remains still an unexplored
area of nanoscience in the sense of nanotechonology.

 Even at basic research level the tailored and controlled way of surface nanofabrication is still challenging
although a great deal of efforts have been done in the last decades towards the understanding
of processes responsible for the formation of regular patterns \cite{sculpting,Facsko,ripples}.

  Ion-sputtering induced self-organized surface nanopatterning is a widely
used buttom-up technique for nanofabrication mostly still at a 
basic research level,
however, possible potential application fields such as the mass-fabrication of nanotemplates remains still challenging  \cite{sculpting,Facsko}.
Although the cost-effective
self-organized regulation of the nanotopography of nanotemplates
is still lacking, the ability of manipulating the surface topography in the nanoscale using a single technological step (or replacing an another top-down step) based on the phenomenon of
slef-assembly is an 
important issue and a great challenge in nanoscience.

 The 
self-organization induced formation of dot pattern has  already been demonstrated on
semiconductors \cite{Facsko}.
Periodic ripple formation on surfaces has been the subject of numerous
experimental \cite{ripples} and  theoretical works \cite{Makeev,Chason}.
The formation of ripples on Si has also been reported recently  \cite{rippleonSi,Facsko09}.
In general, however, the fundamental understanding of the self-organization of nanoscale periodic surface features
is still lacking.

 In order to understand the processes which could controll
self-organized NP an efficient simulation tool is required.
The available theroretical and computational (numerical) techniques mostly do not go beyond the level of
phenomenology and continuum approaches \cite{Heinig}.
Recently, impurity effects have been cosnidered within a continuum model \cite{phenomenFe}.
Using these approaches, however, the atomistic details of NP remain inaccesible.

 Atomistic computer simulations based on kinetic Monte Carlo approach have already been used for
modelling self-organized nanopatterning \cite{Chason,Stepanova,Yewanda,Strobel}.
Although, molecular dynamics (MD) simulations have yet been employed for
following the time evolution of ion-patterning with some success \cite{Sule_2011,Sule_JCP09},
however, simulations starting from a flat surface has not yet been
utilized.
In the present work we would like to show that upon low-fluence
ion-sputtering early-patterning takes place with features
similar to that of occur at higher fluences.
Prepatterns which can be obtained upon $< 10^{16}$ ion/c$m^{-2}$
fluences although show irregular features, the occuring network of adatom islands
could evolve into regular patterns upon higher ion-fluence.
Hence in the early stage of NP a random disorder is found which is, however,
shows the characteristics of nanoporous Si under certain conditions.
The transition of the disordered to ordered topography transition is still unreachable with
the available computational techniques.

 The key questions to be answered are the followings: \\
 How to rule the surface topography ?  \\
 Which steps are encountered during the surface height evolution
 by the topography ? \\
 How various patterns transform into each other ?
 What is the atomistic mechanism of self-organization of 
  regular surface features ? \\

\begin{figure}[hbtp]
\begin{center}
\includegraphics*[height=6cm,width=8cm,angle=0.]{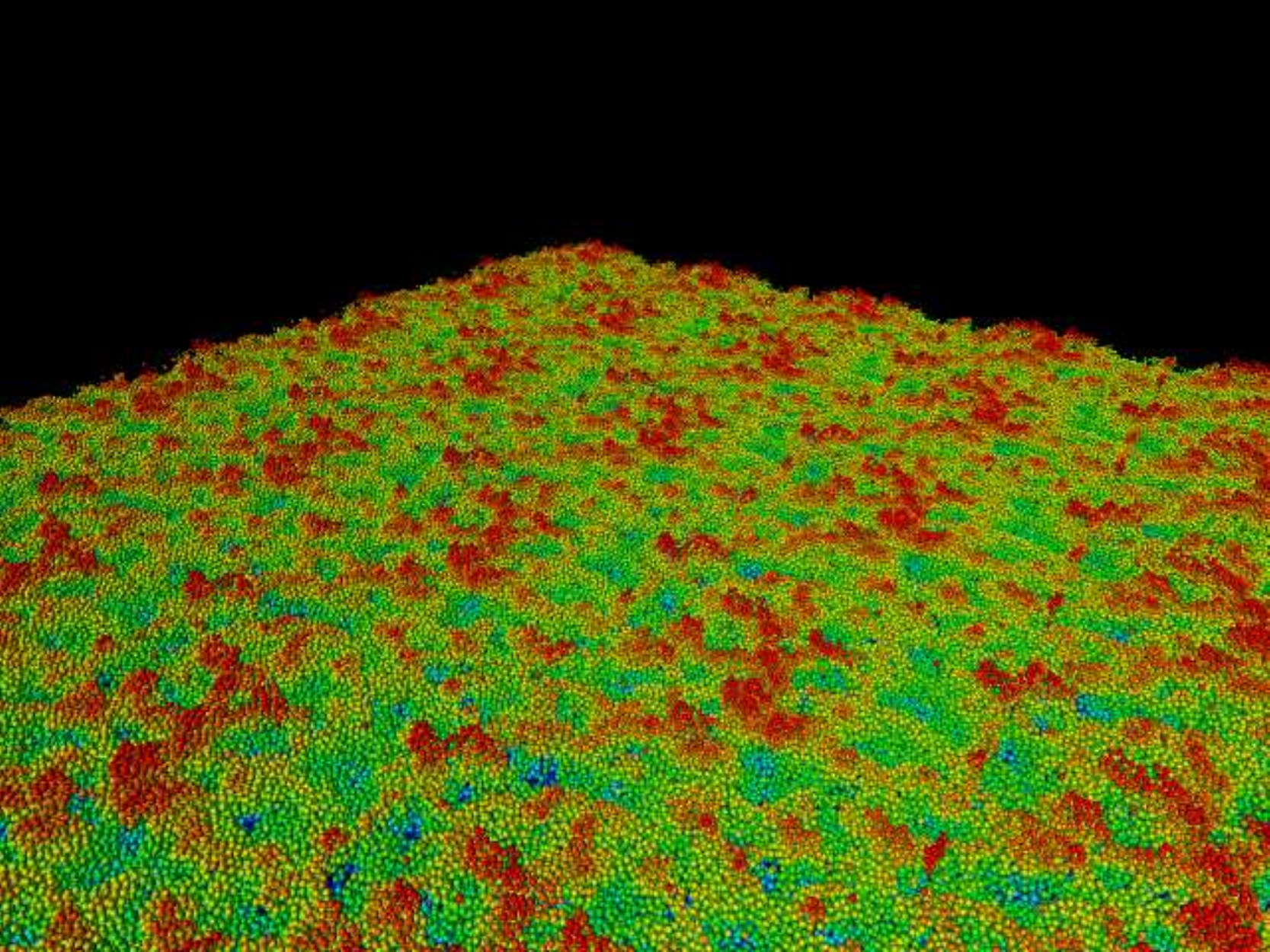}
\includegraphics*[height=6cm,width=8cm,angle=0.]{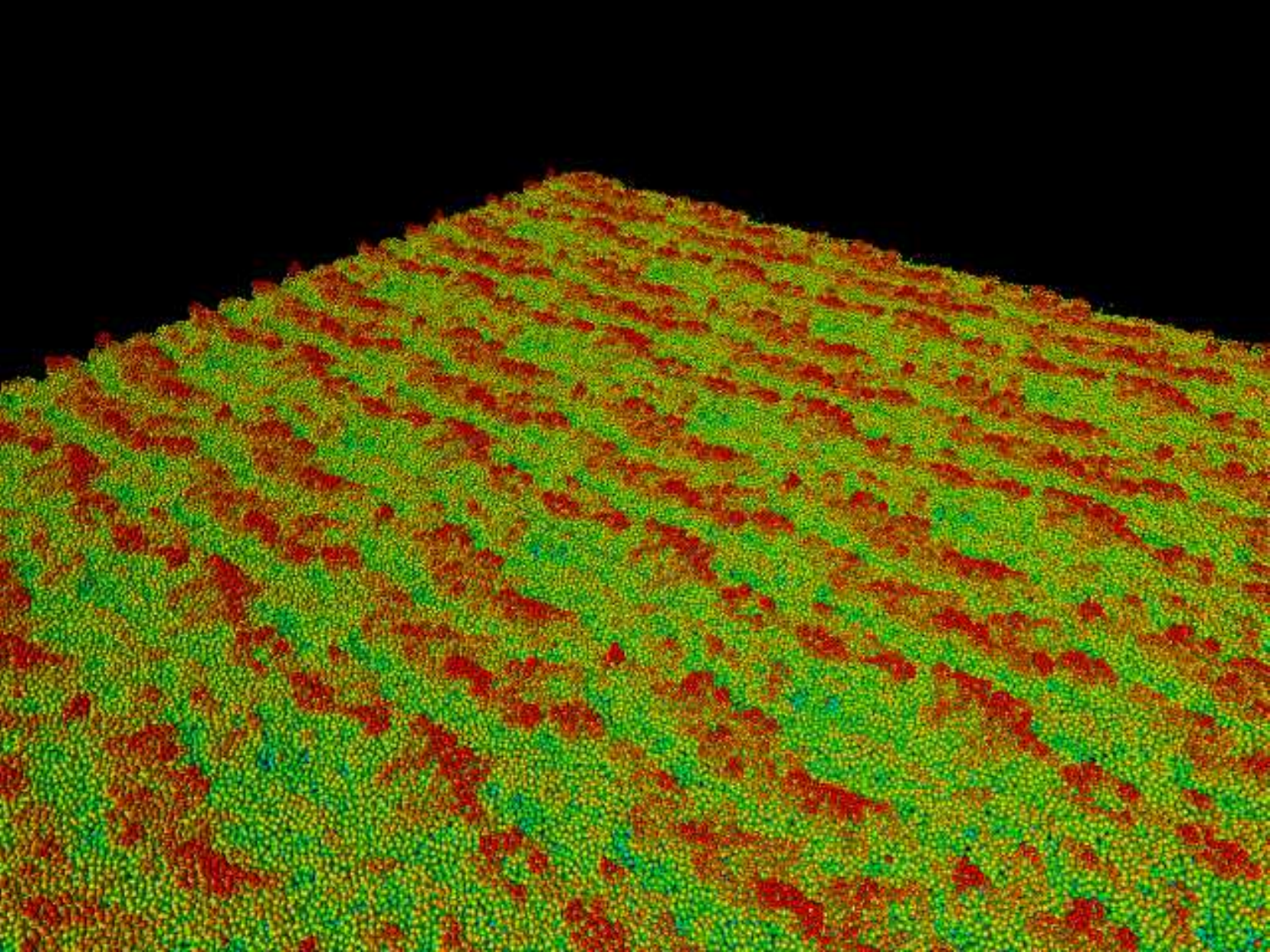}
\caption[]{
The color coded topography images of the bombarded surfaces
with (ibad, Fig 1a ) and without (noibad, Fig 1b) Fe impurities.
}
\label{fig1}
\end{center}
\end{figure}

\begin{figure}[hbtp]
\begin{center}
\includegraphics*[height=6cm,width=8cm,angle=0.]{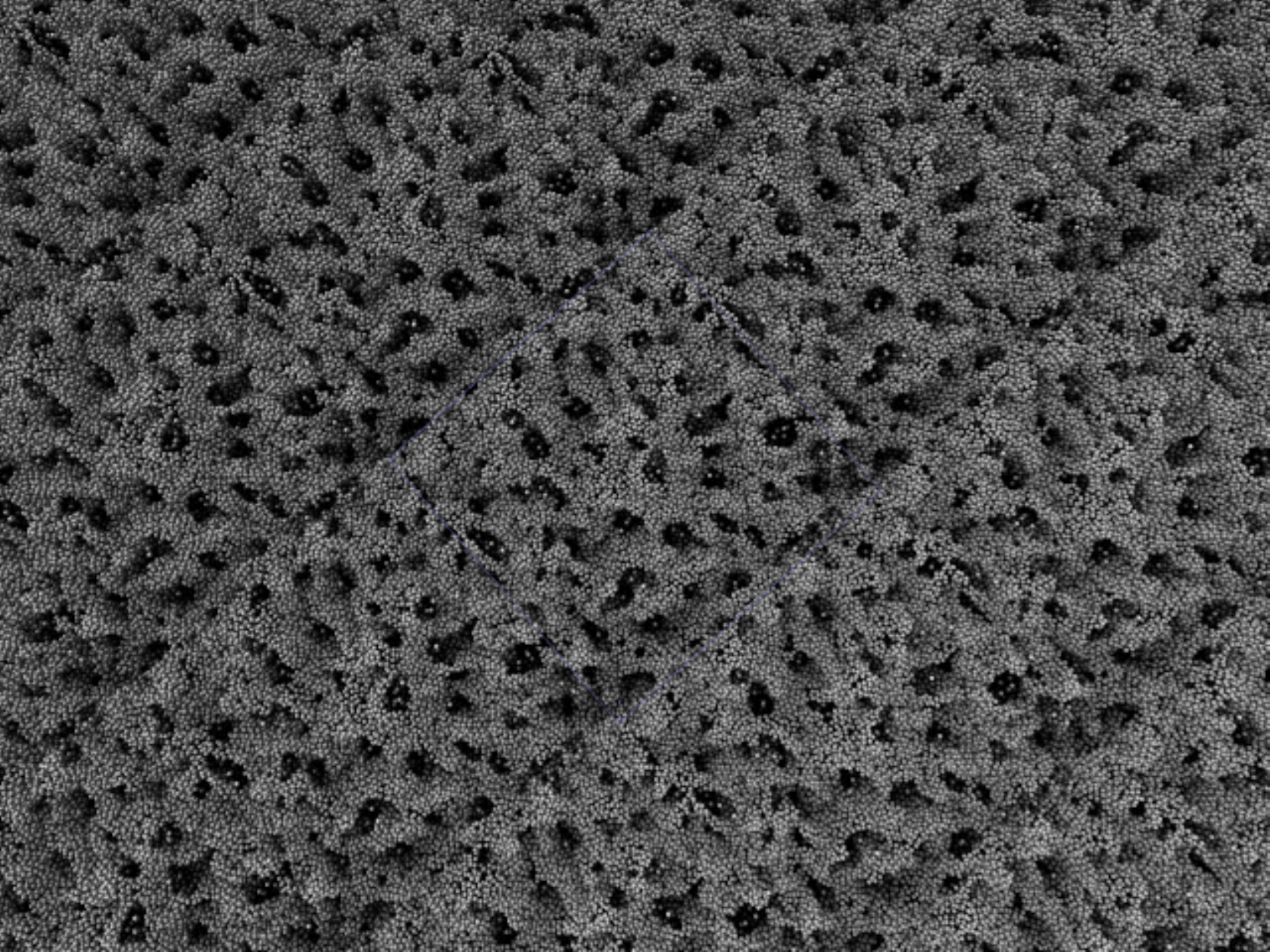}
\includegraphics*[height=6cm,width=8cm,angle=0.]{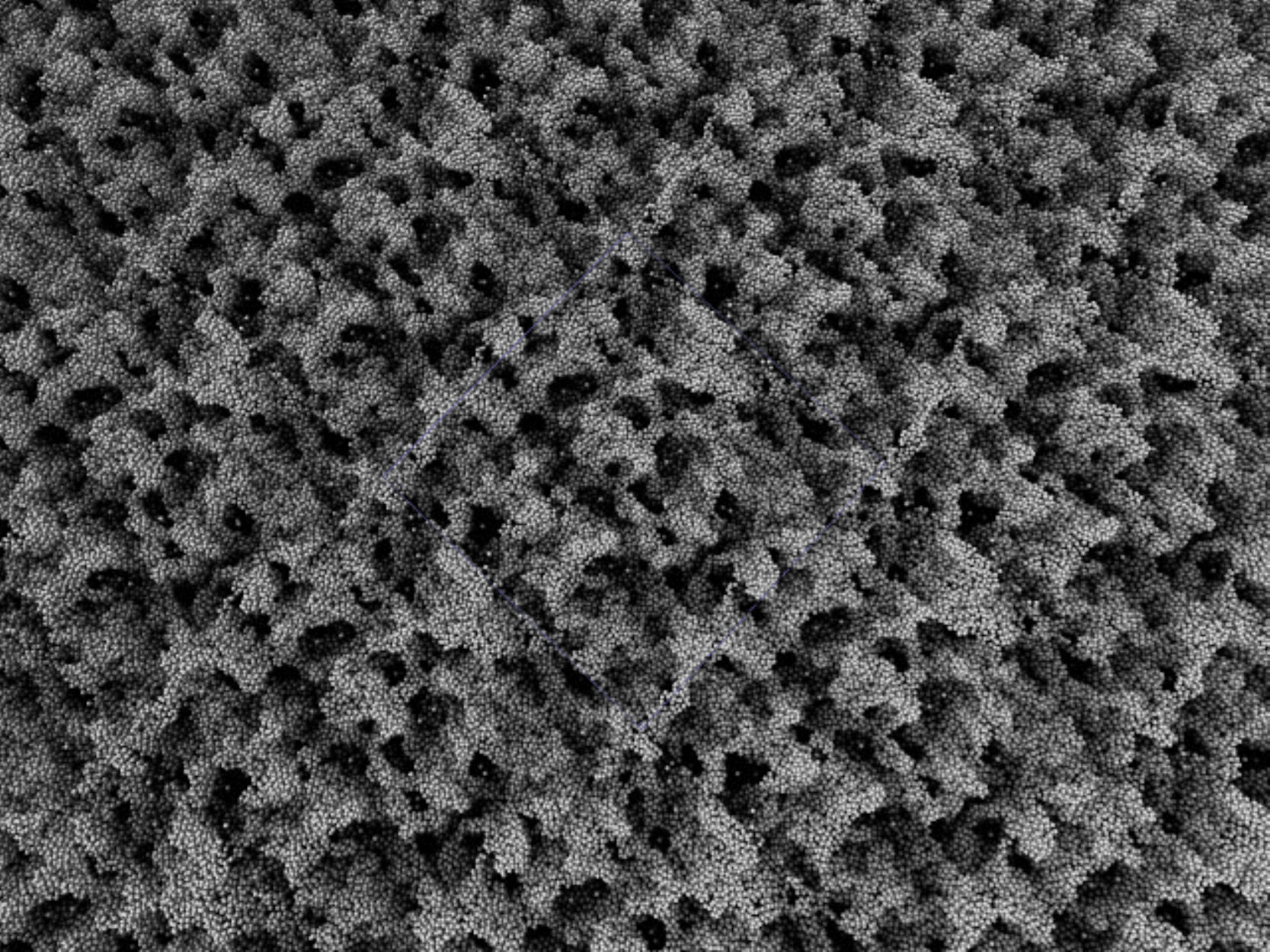}
\includegraphics*[height=6cm,width=8cm,angle=0.]{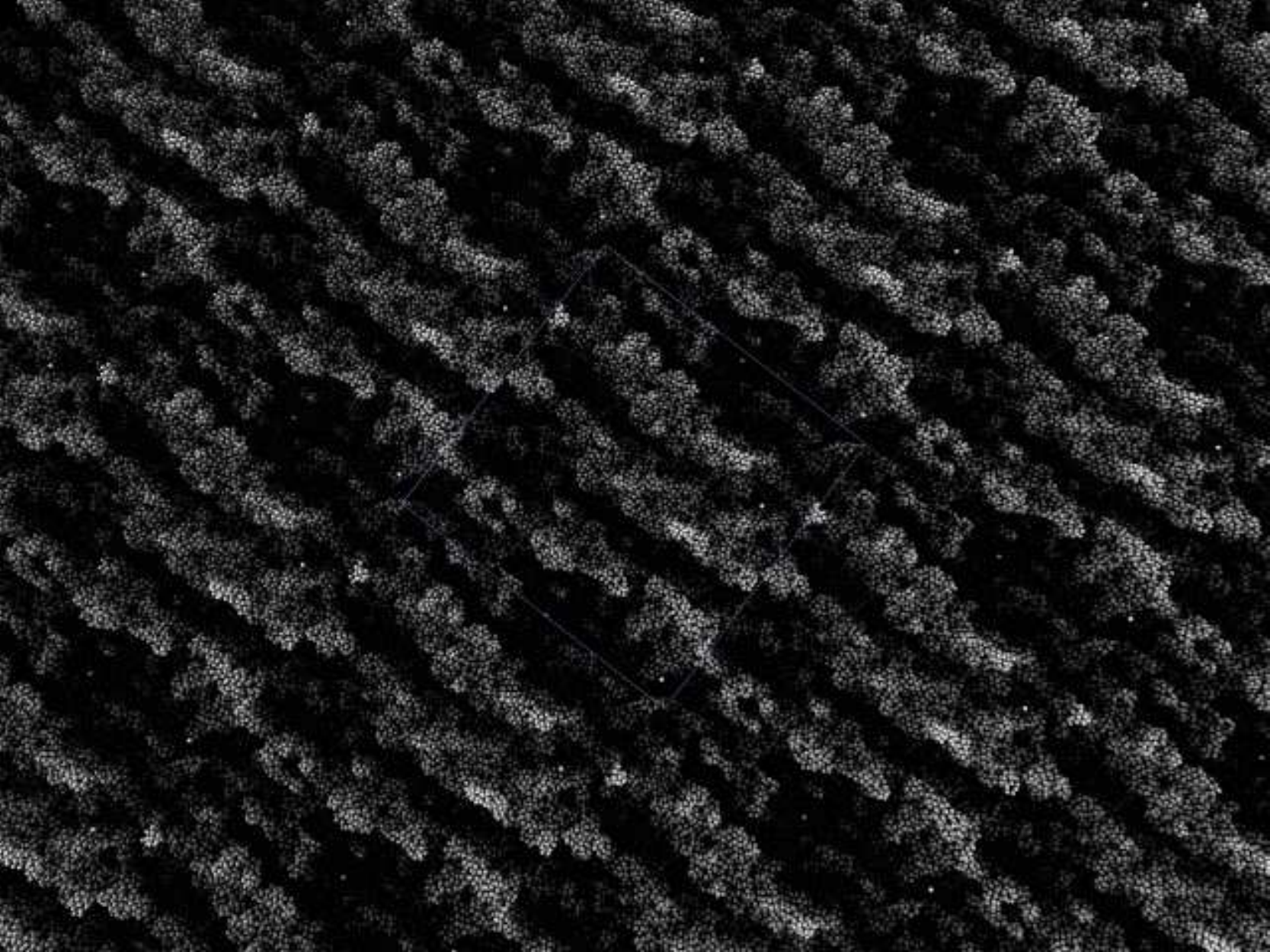}
\caption[]{
Color coded (gray) height images of ion-sputtered surfaces of aSi
(without Fe impurities, no ibad)
at $30^{\circ}$, $50^{\circ}$ and $70^{\circ}$ using $500$ eV $Xe^{+}$ impacts
(~1300 repeated impacts randomly distributed over the whole area).
Periodic images of the original simulation cell is shown magnified
3 times (the original cell size was $162$ $\hbox{\AA}$) and the
present images correspond to roughly $80 \times \sim 80$ nm$^2$ area.
}
\label{xesi}
\end{center}
\end{figure}

\begin{figure}[hbtp]
\begin{center}
\includegraphics*[height=6cm,width=8cm,angle=0.]{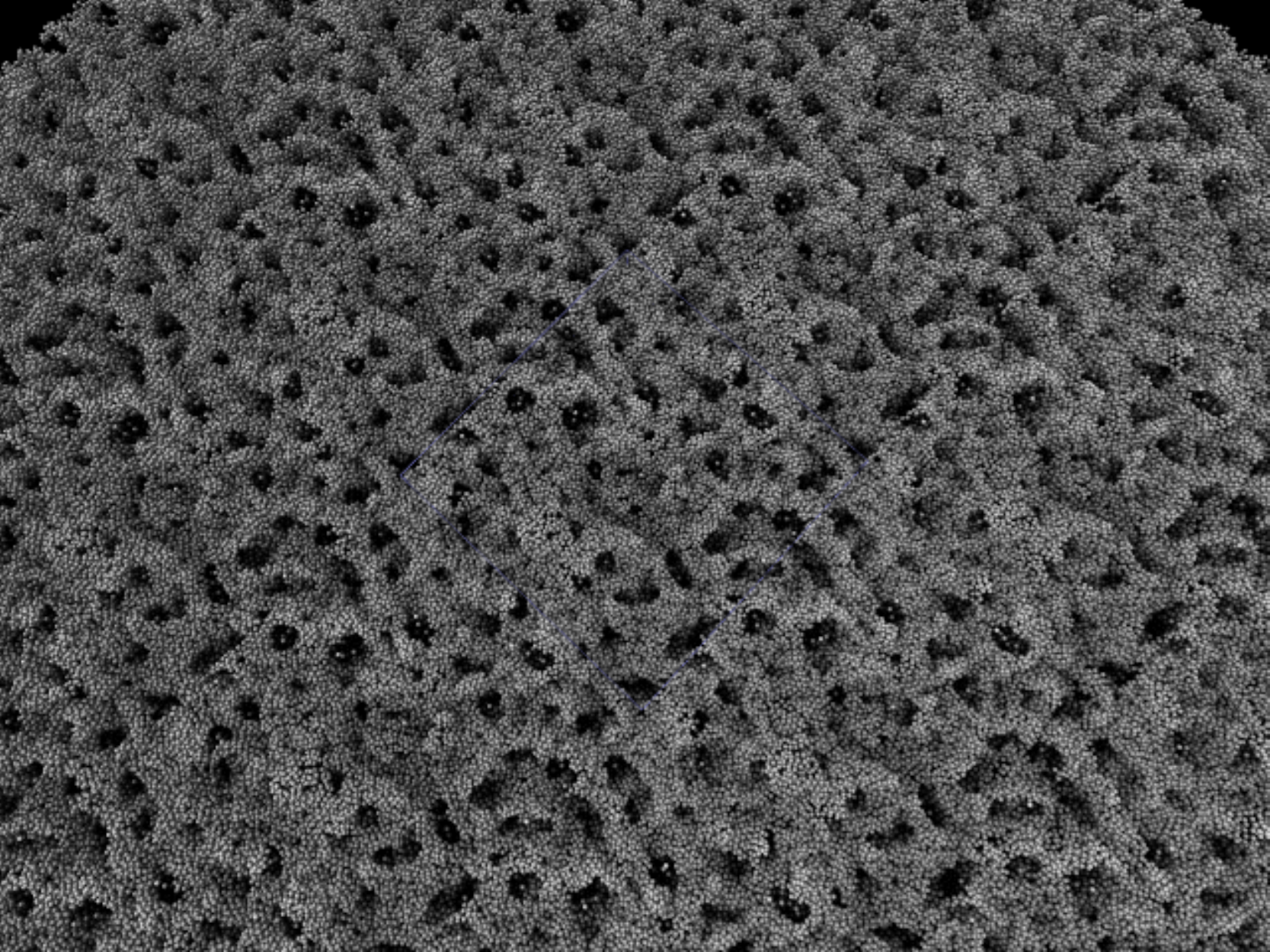}
\includegraphics*[height=6cm,width=8cm,angle=0.]{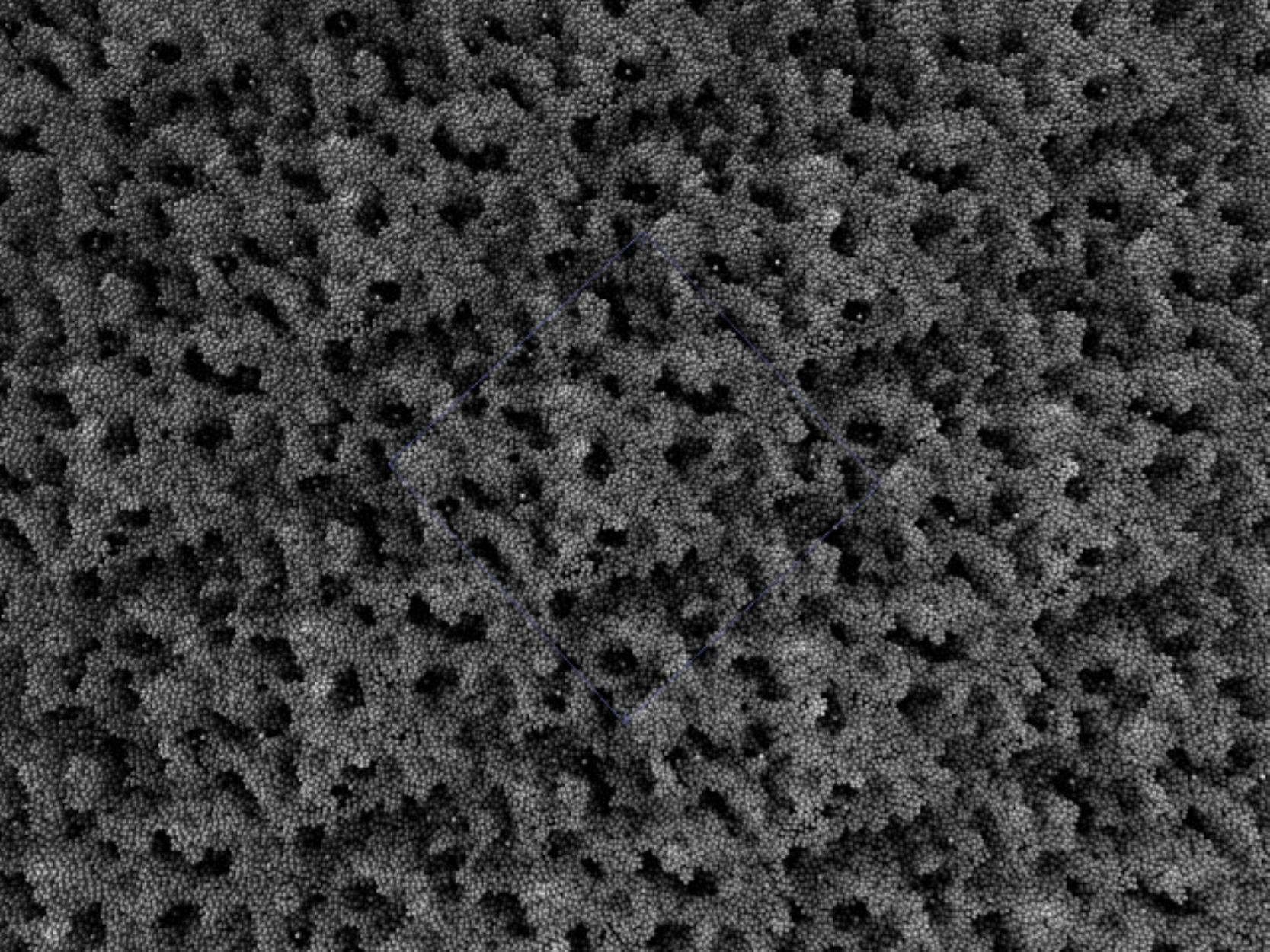}
\includegraphics*[height=6cm,width=8cm,angle=0.]{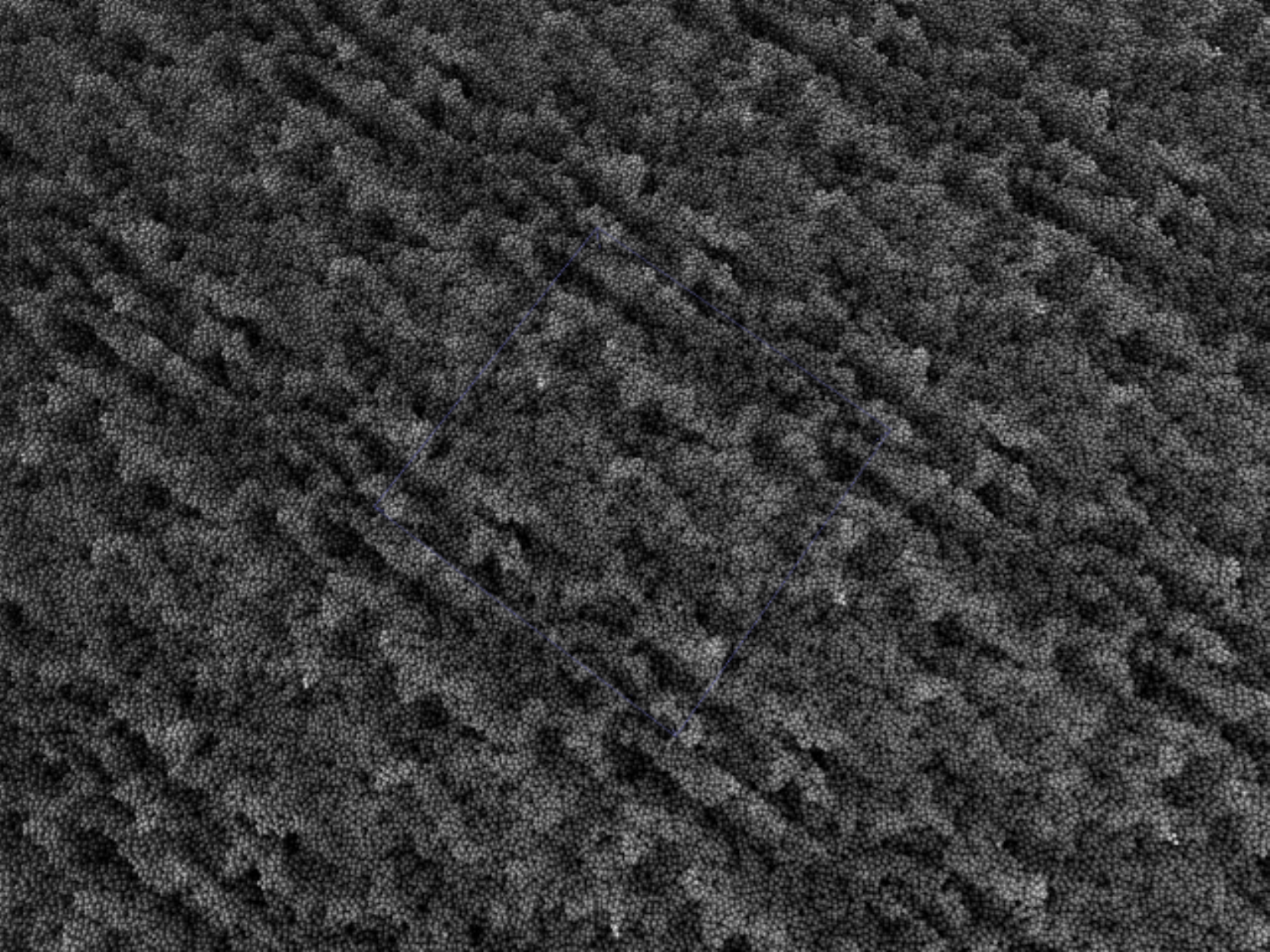}
\caption[]{
Color coded (gray) height images of ion-sputtered surfaces of aSi
(with Fe impurities, ibad)
at $30^{\circ}$, $50^{\circ}$ and $70^{\circ}$ using $500$ eV $Xe^{+}$ impacts
(~1000 repeated impacts randomly distributed over the whole area).
Periodic images of the original simulation cell is shown magnified
3 times (the original cell size was $162$ $\hbox{\AA}$) and the
present images correspond to roughly $80 \times \sim 80$ nm$^2$ area.
}
\label{xesi}
\end{center}
\end{figure}

\begin{figure}[hbtp]
\begin{center}
\includegraphics*[height=6cm,width=8cm,angle=0.]{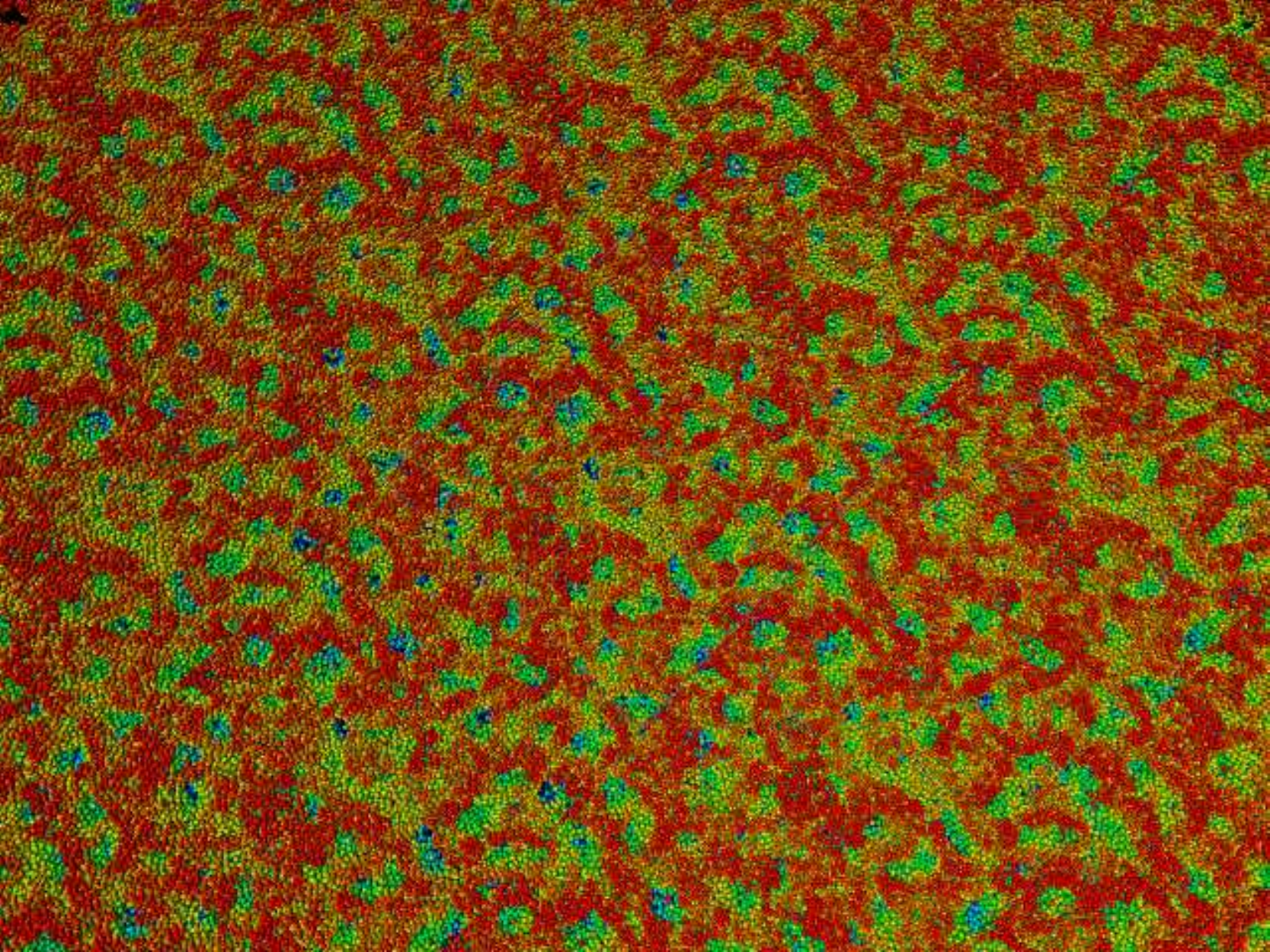}
\includegraphics*[height=6cm,width=8cm,angle=0.]{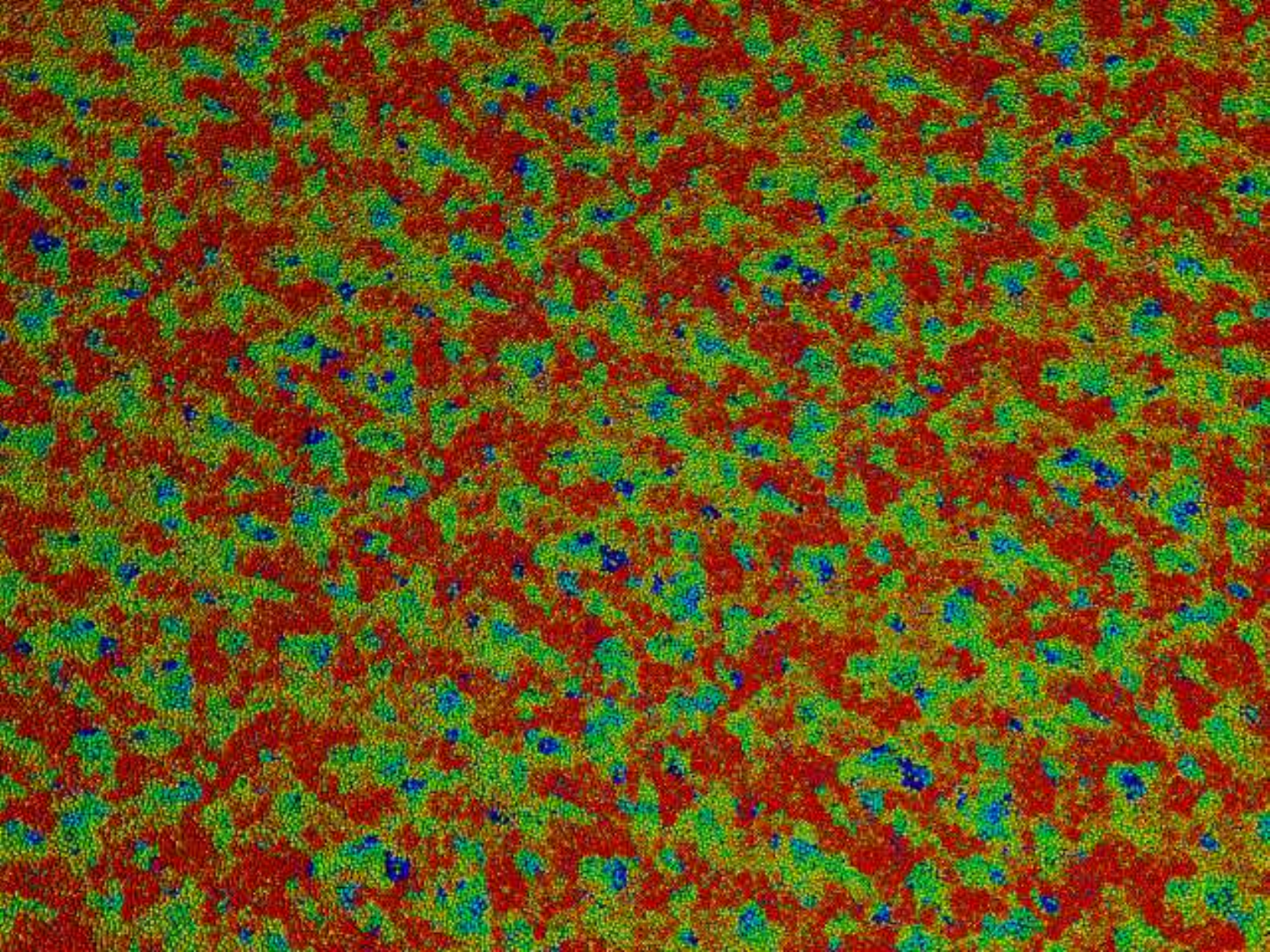}
\includegraphics*[height=6cm,width=8cm,angle=0.]{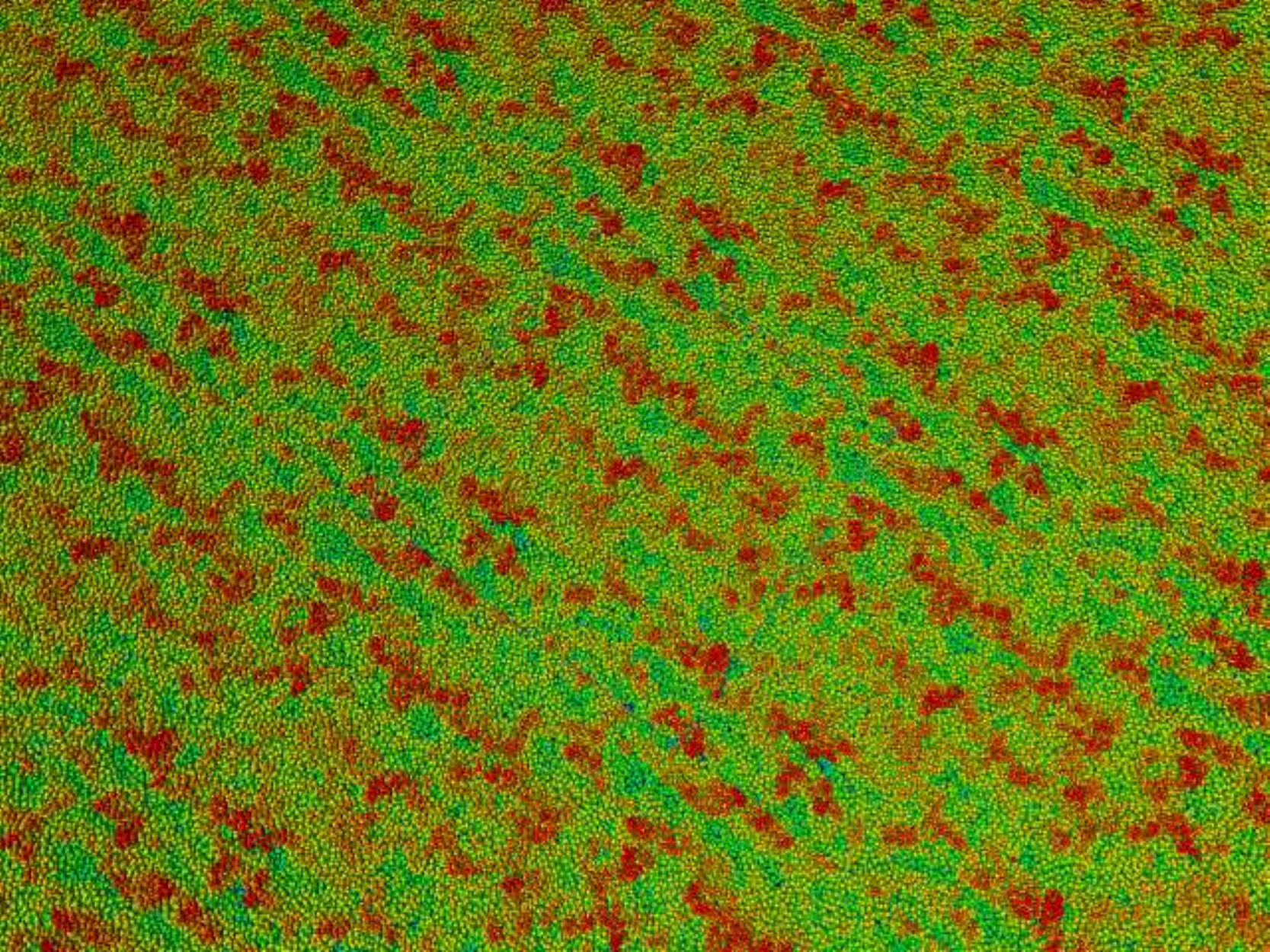}
\caption[]{
Color coded (rainbow) height images of ion-sputtered surfaces of aSi
(with Fe impurities, ibad)
at $30^{\circ}$, $50^{\circ}$ and $70^{\circ}$ using $500$ eV $Xe^{+}$ impacts
(~1000 repeated impacts randomly distributed over the whole area).
Periodic images of the original simulation cell is shown magnified
3 times (the original cell size was $162$ $\hbox{\AA}$) and the
present images correspond to roughly $80 \times \sim 80$ nm$^2$ area.
}
\label{xesi}
\end{center}
\end{figure}


\begin{figure}[hbtp]
\begin{center}
\includegraphics*[height=6cm,width=8cm,angle=0.]{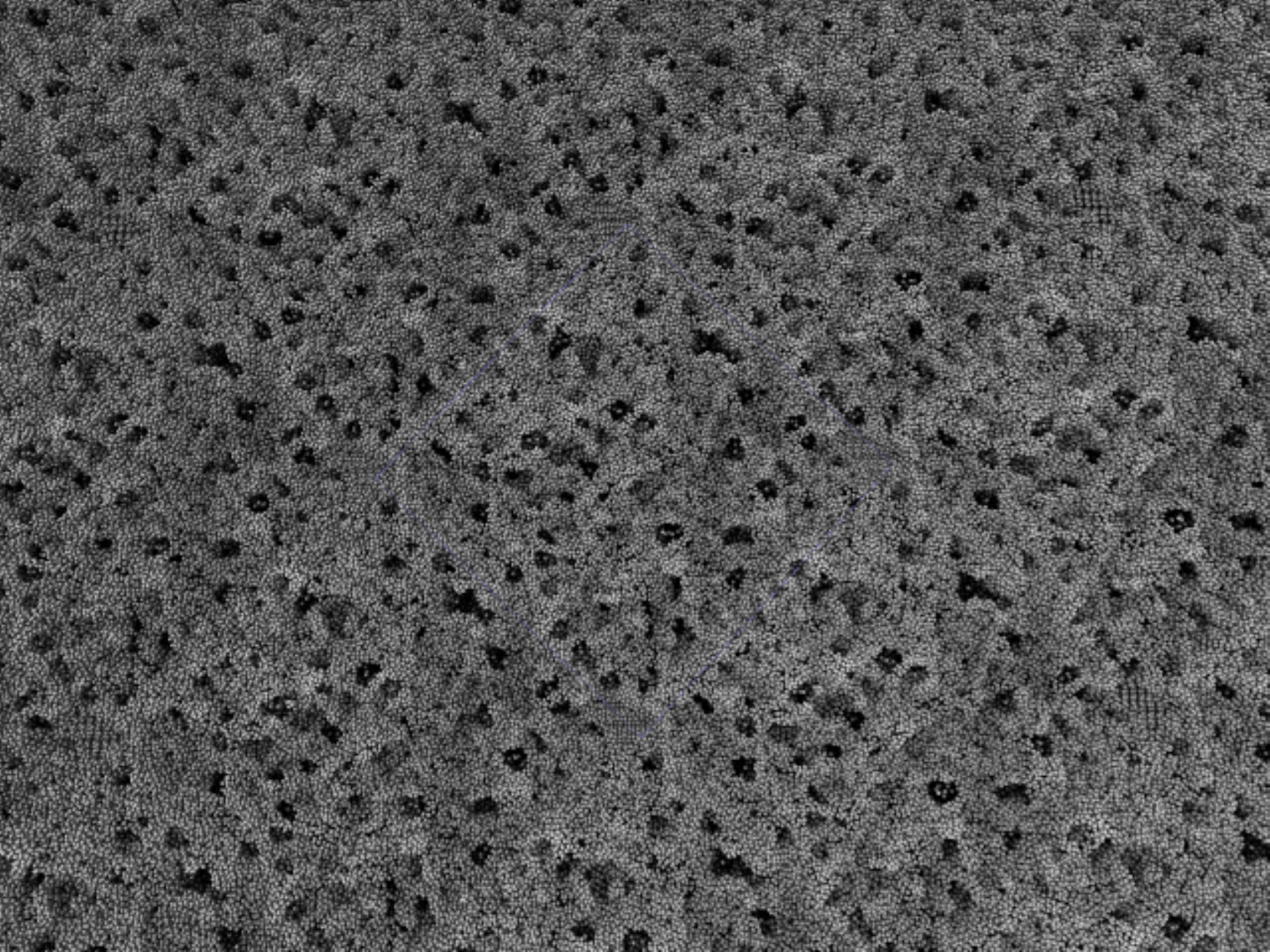}
\includegraphics*[height=6cm,width=8cm,angle=0.]{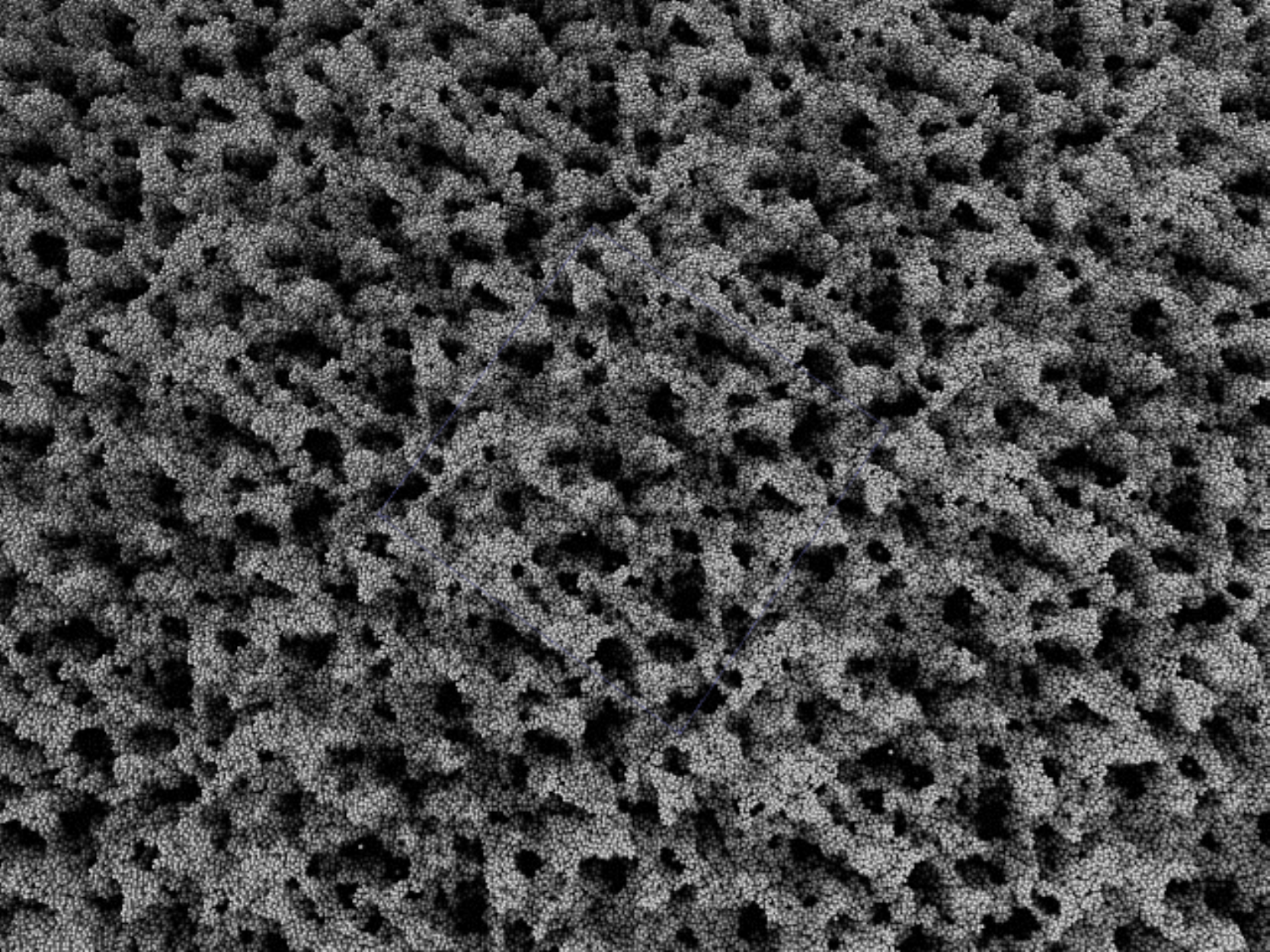}
\includegraphics*[height=6cm,width=8cm,angle=0.]{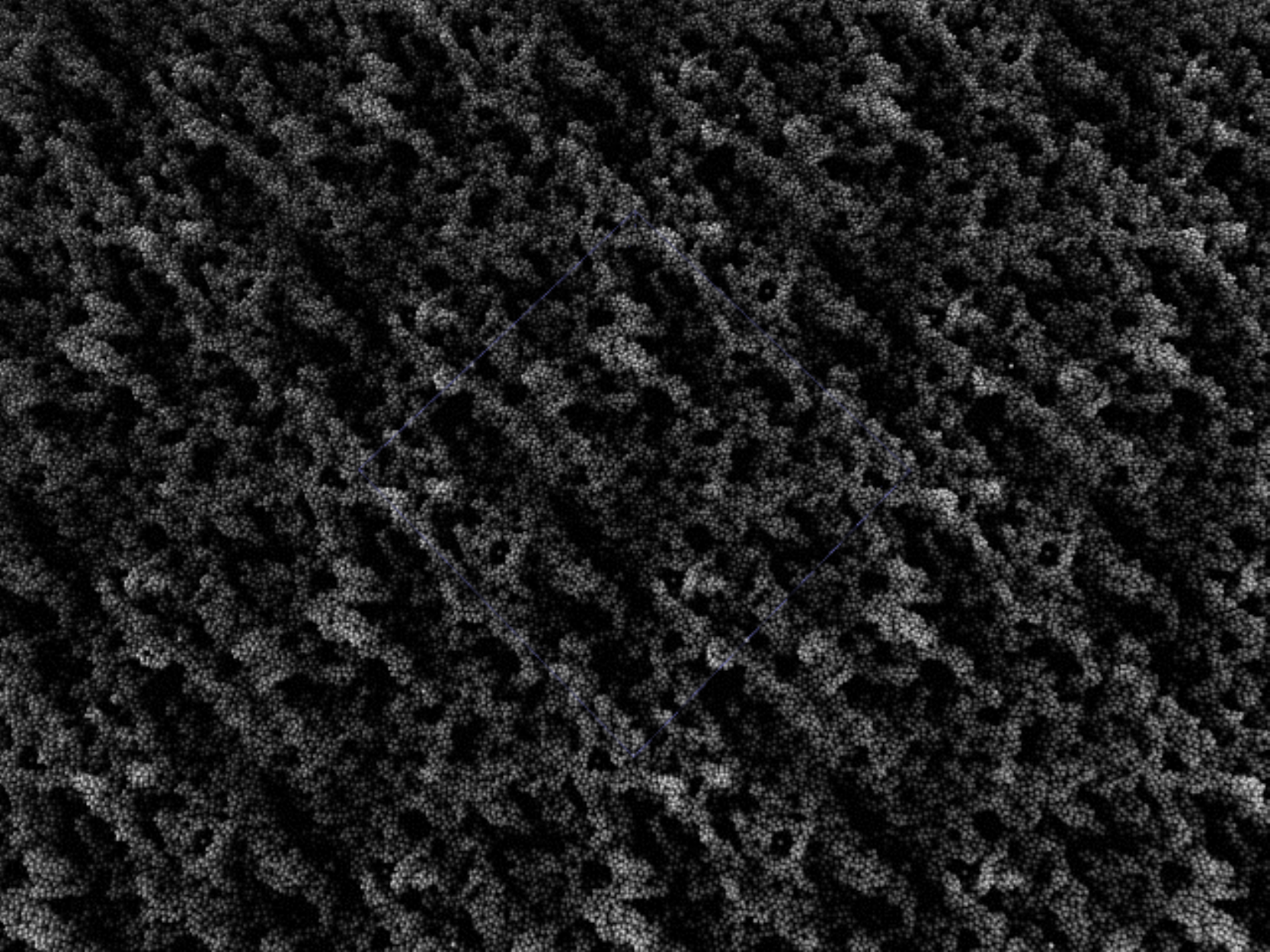}
\caption[]{
Color coded (gray) height images of ion-sputtered surfaces of Si(001)
codeposited with Fe (ibad)
$30^{\circ}$, $50^{\circ}$ and $70^{\circ}$ using $500$ eV $Kr^{+}$ impacts
(~1000 repeated impacts randomly distributed over the whole area).
}
\label{xesi}
\end{center}
\end{figure}

\begin{figure}[hbtp]
\begin{center}
\includegraphics*[height=6cm,width=8cm,angle=0.]{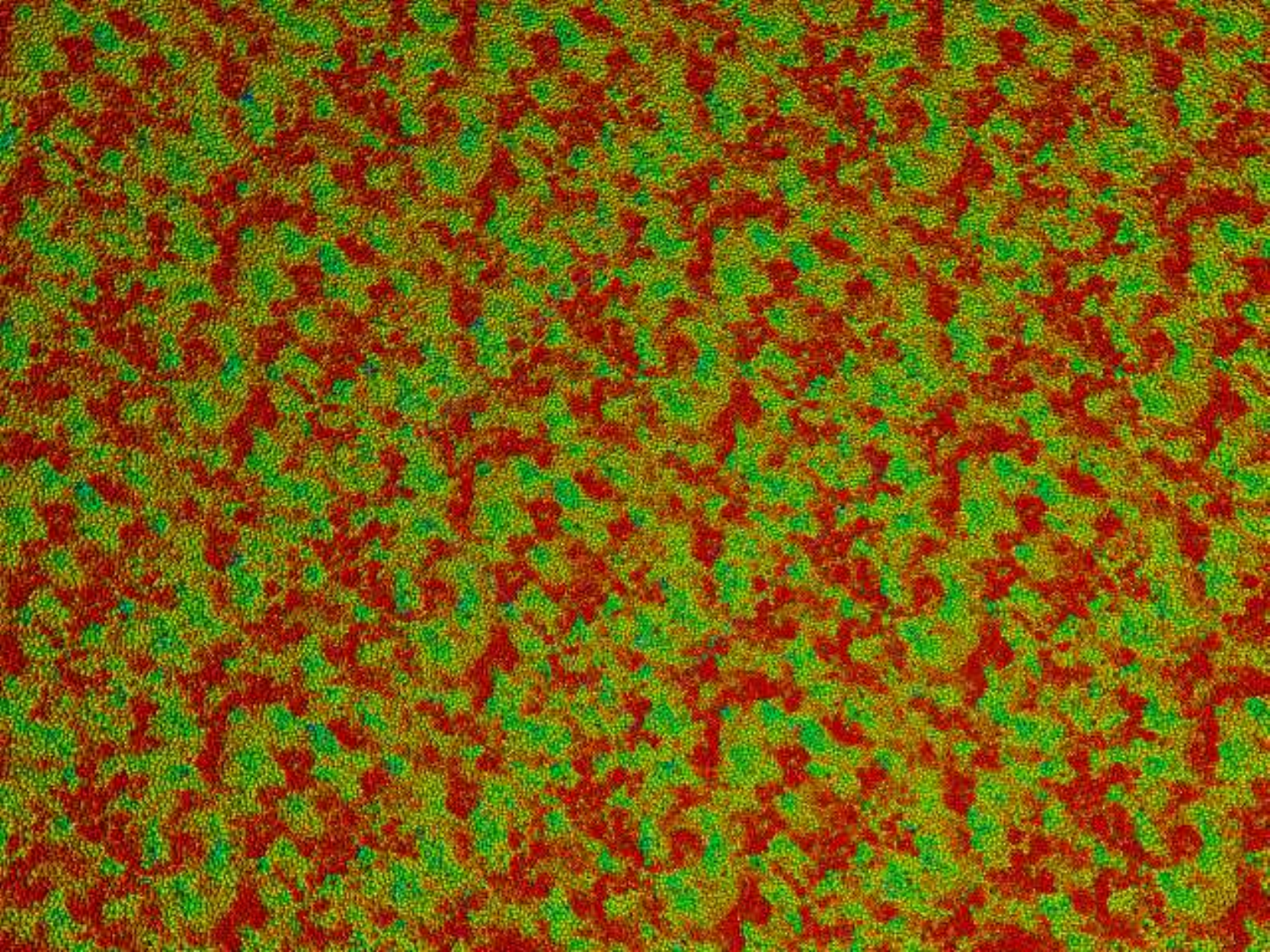}
\includegraphics*[height=6cm,width=8cm,angle=0.]{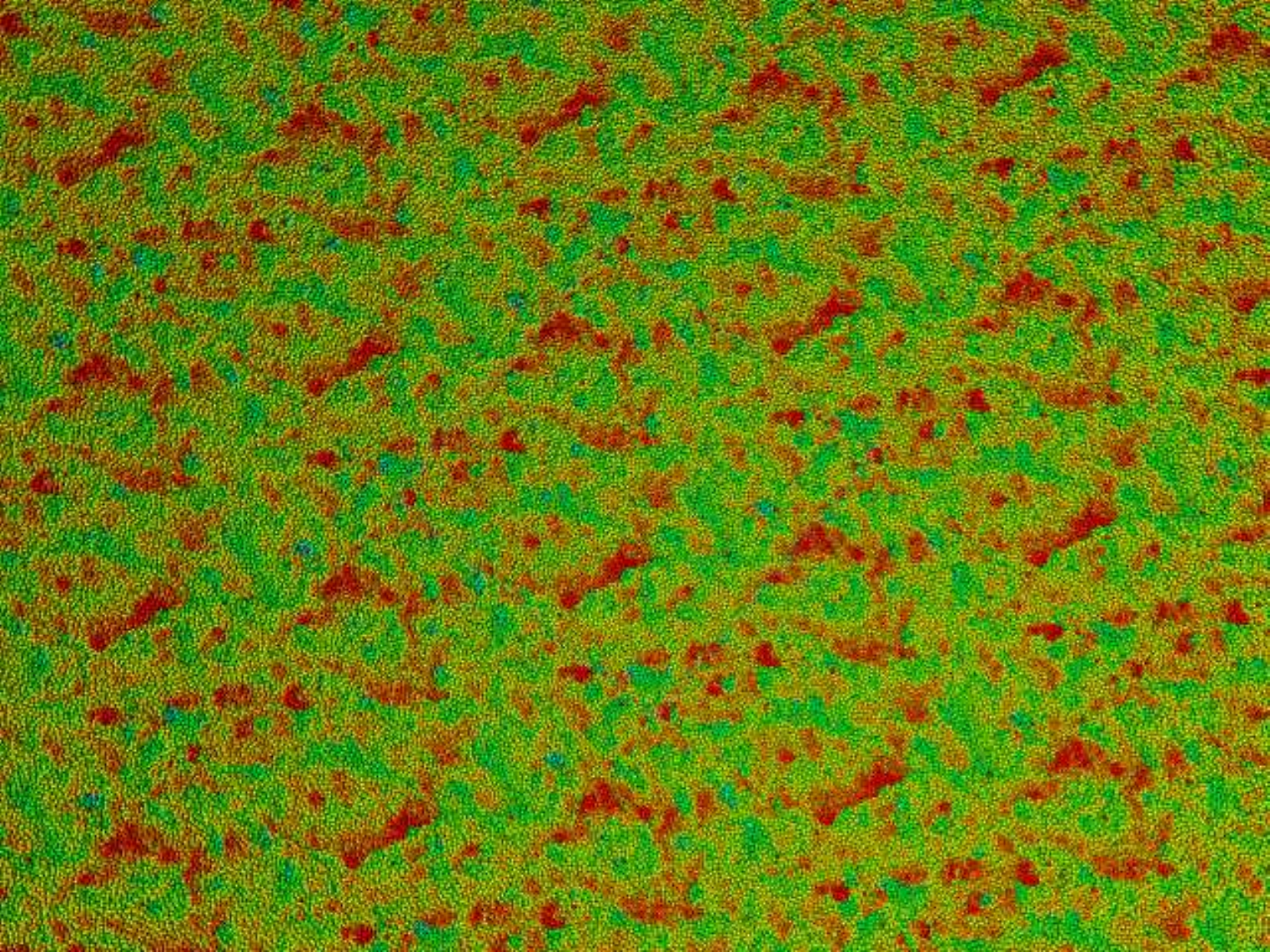}
\caption[]{
Color coded (gray) height images of ion-sputtered surfaces of FeSi
without and with Fe codeposition (Fig 1a and 1b)
at $50^{\circ}$ $500$ eV $Xe^{+}$ impacts
(~1000 repeated impacts randomly distributed over the whole area).
}
\label{xesi}
\end{center}
\end{figure}


\section{The fitting of the Albe-Erhart empirical potential for FeSi}





It is assumed that the energy of a system can be separated into a sum of
pairwise contributions $V_{ij}$,
\be
E=\sum_{ij,i>j} f_{ij}(r_{ij})[V_{ij}^R-b_{ij} V^A_{ij}(r_{ij})],
\ee
where for the repulsive and attractive pair interactions
we have the following formulas :
\be
V_{ij}^R=\frac{D_0}{S-1} exp(-\beta \sqrt{2S}(r-r_0)),
\ee
\be
V_{ij}^A=\frac{SD_0}{S-1} exp(-\beta \sqrt{2/S}(r-r_0)),
\ee
respectively.

   The bond-order $b_{ij}$ depends on the chemical environment of the atoms i and j. It is given
by
\be
b_{ij}=(1+\chi_{ij}^n)^{\frac{1}{2n}}
\ee
with
\be
\chi_{ij}=\sum_{k(\neq i,j)} f_{ik}^c(r_{ik}) g_{ik}(\Theta_{ijk})  exp[2\mu_{ik}(r_{ij}-r_{ik})]
\ee
and where the angular term $g(\Theta)$,
\be 
g(\Theta)=\gamma \biggm(1+\frac{c^2}{d^2}-\frac{c^2}{d^2+(h+cos\Theta)^2}\biggm).
\ee
   By construction, the BOP potential includes only nearest neighbor interactions. This
necessitates the use of a cutoff function $f_{ij}(r)$, which scales the energy between the first
and second neighbors shells smoothly to zero.

The cutoff function is
\[
f_{ij}(r_{ij})= \left\{ \begin{array}{cc}
~~~~~~~~~ 1 & r \le R_c-D_c  \\ \frac{1}{2}-\frac{1}{2}sin[\frac{\pi}{2}(r-R_c)/D_c] & |r-R_c| \le D_c
 \\
0 & r \ge R_c+D_c
\end{array} \right. \]
where $R_c$ $\hbox{\AA}$ is the cutoff distance
and $D_c$ $\hbox{\AA}$ is the damping distance.

The BOP potential has been parametrized using an extended training set of
various structures (B1 (NaCl), B2 (CsCl), B3 (ZnS) and B20 (eps-FeSi) phases for FeSi).
The parametrization procedure has been carried out using the PONTIFIX code
developed by P. Erhart \cite{pontifix}.
The parameters of FeSi have simulataneously been fitted together
with the elemental Si and Fe parameters.
For the elements similar data base has been used given by
Albe and Erhart \cite{pontifix}.
{\em ab initio} SIESTA \cite{SIESTA} calculations were used to determined the cohesive energies
of various structures.
   A Levenberg-Marquardt least-squares algorithm \cite{Press} has been implemented
in Pontifix to find a combination of parameters which minimizes the deviation
between the properties in the fitting database and the properties predicted by the potential.
Parameter sets for different interaction types can be fitted simultaneously.
The fitting database encompassed the bond lengths and energies of various structures as well as elastic
constants.


\begin{table}[t]
\caption[]
{
The fitted parameters used in the bond order interatomic potential for
Fe-Si, Fe-Fe and for Si-Si.
}
\begin{ruledtabular}
\begin{tabular}{cccc}
 & \bf{Fe-Si} & \bf{Fe-Fe} & \bf{Si-Si}  \\
\hline
$D_0$ (eV)                   & 4.635031 & 2.877707 & 3.761432 \\
$r_0$ ($\hbox{\AA}$)         & 1.546801 & 1.595041 & 2.170624 \\
S                          & 2.330520 & 8.385355 & 1.536018 \\
$\beta$ ($\hbox{\AA}^{-1}$)  & 1.058204 & 1.127692 & 1.285636 \\
$\gamma$                     & 0.078319 & 0.026045 & 0.099010 \\
c                          & 0.403341 & 1.481643 & 0.350297 \\
d                          & 0.197786 & 0.234050 & 0.280854 \\
h                         & -0.181581 & 0.263854 & 0.220250 \\
n                          & 0.782953 & 5.417789 & 0.955404 \\
2 $\mu$                      & 0.0      & 0.0      & 0.0      \\
$R_c$ ($\hbox{\AA}$)         & 2.983063 & 2.444649 & 3.299024 \\ 
$D_c$ ($\hbox{\AA}$)         & 0.2      & 0.2      & 0.15  \\
\end{tabular}
\end{ruledtabular}
\footnotetext[1]{
}
\label{T1}
\end{table}

\begin{table}[t]
\caption[]
{
The test of the preformance of the BOP potential with the new parameter set
}
\begin{ruledtabular}
\begin{tabular}{ccc}
 &   present work & experimental    \\
\hline
\bf{FeSi} (B20) &   &  \\ 
\hline
lattice constant ($\hbox{\AA}$) &  4.52 &  4.49  \\
melting point (K)               &  1600 $\pm$ 100 & 1550    \\
cohesive energy (eV/atom)       &  4.89 &  4.87 \\ 
surface energy (eV/atom)        &  0.59 & 0.8  \\
bulk modulus (GPa)              &  185. & 160. \\
\hline
\bf{Fe} & &  \\ 
\hline
lattice constant ($\hbox{\AA}$) &  2.83 &  2.87  \\
melting point (K)               &  1900 $\pm$ 50 & 1810   \\
cohesive energy (eV/atom)       &  4.31 &  4.28   \\
surface energy (eV/atom)        &   &   \\
bulk modulus (GPa)              &  129 & 169   \\
 B'                             &  3.1 & 5.1  \\
\hline
\bf{Si} & &  \\
\hline
lattice constant ($\hbox{\AA}$) &  5.43 & 5.43  \\
melting point (K)               &  1730 $\pm$ 50 & 1687   \\
cohesive energy (eV/atom)       &  4.63 &  4.63   \\
surface energy (eV/atom)        &  0.8-1.7 & 1.2  \\
bulk modulus (GPa)              &  102 & 99   \\
\end{tabular}
\end{ruledtabular}
\label{T1}
\end{table}


\section{The setup of the simulation}

 Classical constant volume molecular dynamics simulations were used to simulate the ion-solid interaction
using the PARCAS code \cite{Nordlund_ref}
and a shell script code written in our laboratory for simulating
ion-sputtering using an atomic serial addition procedure \cite{Sule_JCP09}.  
Further details are given in recent
communications \cite{Sule_JAP07,Sule_JCP09}.
The Tersoff potential has been used for Si together with
a ZBL like repulsive potential which smoothly joined together \cite{tersoff}.
 Comparative studies on the crater formation on Si showed
the reasonable performance for this potential \cite{Si_potentials_crater}.
 A Tersoff-type crosspotentials Xe-Si, Xe-Fe, Kr-Si, Kr-Fe, have been fitted to {\em ab initio} calculations \cite{G03}.
The details of such a fitting process is given elsewhere \cite{Sule_JCP09}.
The profiles of the fitted potentials are shown in Fig. ~\ref{xesi}.

\begin{table}[t]
\caption[]
{
The fitted semiempirical parameters used in the Tersoff interatomic potential for
Xe-Si Kr-Si, Xe-Xe and for Kr-Kr.
}
\begin{ruledtabular}
\begin{tabular}{ccccc}
& A & B & $\lambda$ & $\mu$  \\
\hline
 Xe-Si  & 163.123  & -3291.89  & 125.676 & 3.11012 \\
 Xe-Xe  & 6000.8d0 &  95.2 & 2.47990 & 1.7322   \\
 Kr-Si  & 3723.32  & -750.317 & 5.17001 & 2.94665  \\
 Kr-Kr  & 1082.06  & -715.546 & 2.84947 & 2.85043  \\
\end{tabular}
\end{ruledtabular}
\footnotetext[1]{
Note, that the rest of the parameters of the Tersoff formalism
have not been used which occur in the bond-angle dependent part
by setting $\beta=0.$ \cite{tersoff}.
The Tersoff parameters can be converted into the parameters of
the BOP formalism.
Hereby we use the parameters of the original radial part of the Tersoff 
potential and which has been fitted to ab initio calculations \cite{Sule_JCP09}.
}
\label{T1}
\end{table}

The parameters for the fitted pair-potentials are given in Table ~\ref{T1}.

 The emerging excess heat has been controlled by
the Berendsen heat bath at the bottom of the cells.
 We irradiate the surface of Si(110) in an initially diamond crystal structure
with different length.

 We employ our recently developed computer code for handling conditions
appear during simulated ion-sputtering \cite{Sule_JCP09}.
This code has also seriously been developed and adapted for this particular problem.
Repeated ion impacts
with $0.5$ keV Xe$^+$ ions are initialized  with a time interval of 0.5-5 ps between each of
the ion-impacts. 
The initial velocity direction of the
impacting atom was $30^{\circ}$, $50^{\circ}$ and $70^{\circ}$ with respect to the surface normal.
  We randomly varied the impact position.
In order to approach the real sputtering limit a large number of ion irradiation are
employed using automated simulations conducted subsequently together with analyzing
the history files (movie files) in each irradiation steps.
In this article we present results up to $1000$ ion irradiation which we find suitable for
comparing with low to medium fluence experiments. $1000$ ions are randomly distributed
over a $\sim \lambda \times 54$ \hbox{\AA}$^2$ area which corresponds to $< 10^{15}$
ion/cm$^2$ ion fluence.

\begin{figure}[hbtp]
\begin{center}
\includegraphics*[height=6cm,width=7.5cm,angle=0.]{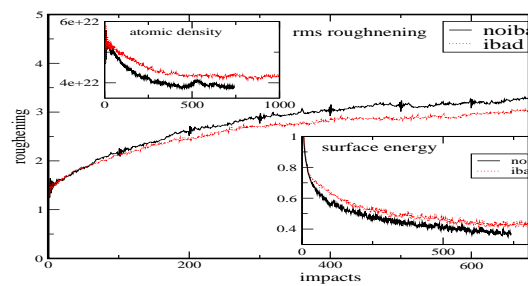}
\caption[]{
The evolution of the surface energy (eV/atom) and
rms surface roughening (Inset) as a function of the Xe$^{+}$ ion impacts
at 500 eV and $30^{\circ}$ incidence angle
for iob-beam assisted deposition and without Fe impurities.
}
\label{surface_energy}
\end{center}
\end{figure}

\begin{figure}[hbtp]
\begin{center}
\includegraphics*[height=6cm,width=7.5cm,angle=0.]{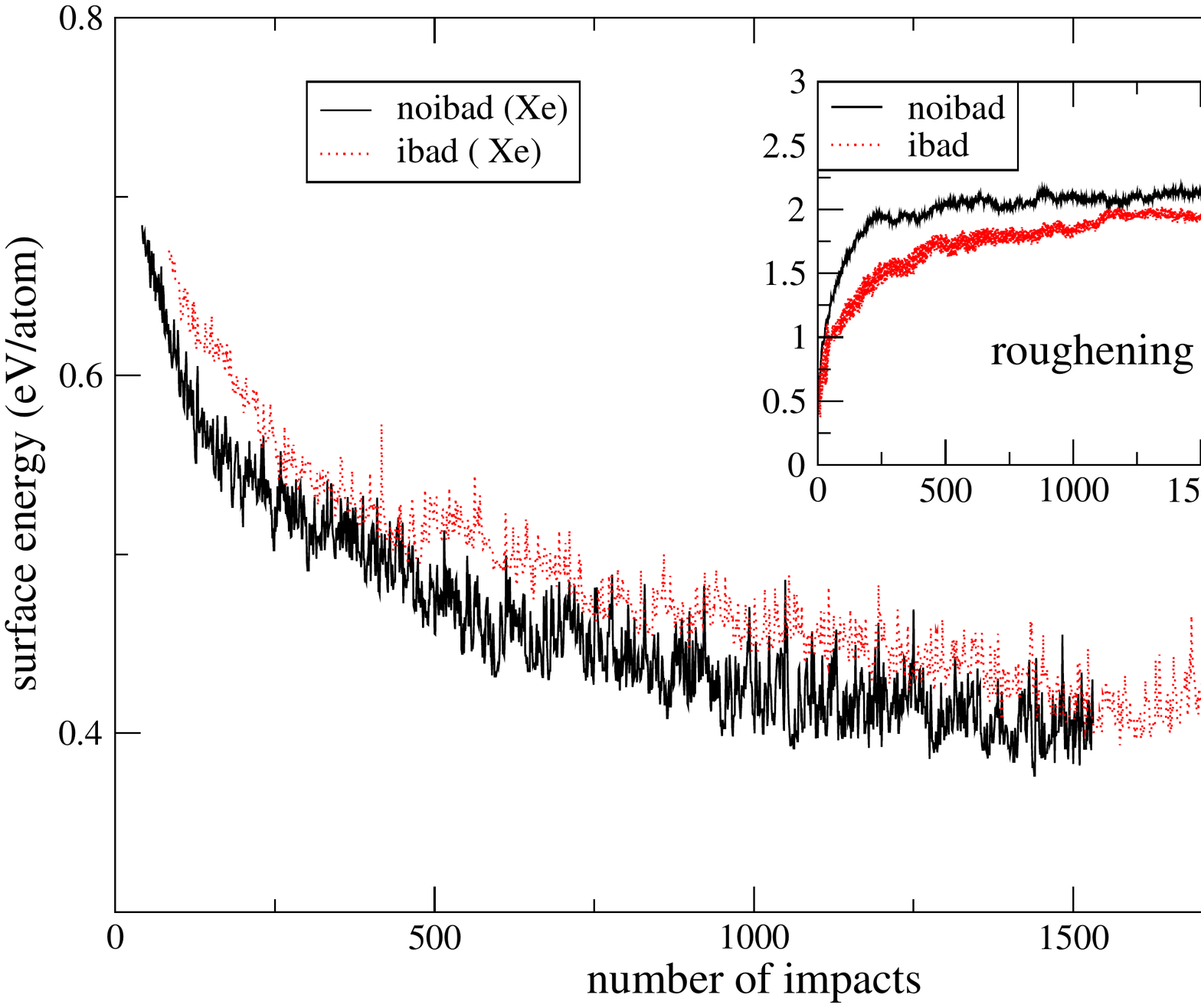}
\caption[]{
The evolution of the surface energy (eV/atom) and
rms surface roughening (Inset) as a function of the Xe$^{+}$ ion impacts
at 500 eV and $50^{\circ}$ incidence angle
for iob-beam assisted deposition and without Fe impurities.
}
\label{surface_energy}
\end{center}
\end{figure}

\begin{figure}[hbtp]
\begin{center}
\includegraphics*[height=6cm,width=7.5cm,angle=0.]{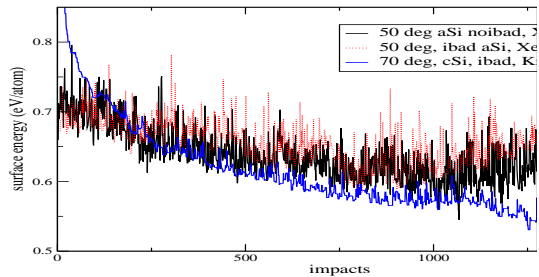}
\caption[]{
The evolution of the surface energy (eV/atom) 
as a function of the Xe$^{+}$ ion impacts
at 500 eV and $50^{\circ}$ and $70^{\circ}$ incidence angle
for iob-beam assisted deposition and without Fe impurities
on aSi and cSi flat surfaces.
}
\label{surface_energy}
\end{center}
\end{figure}

\begin{figure}[hbtp]
\begin{center}
\includegraphics*[height=6cm,width=7.5cm,angle=0.]{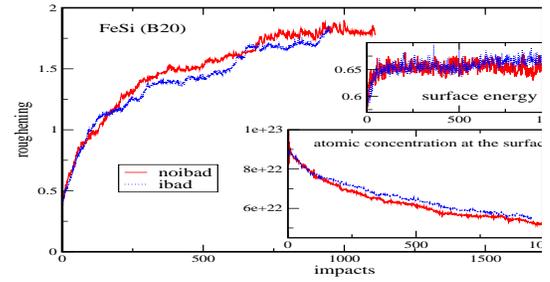}
\caption[]{
The evolution of rms surface roughening, surface energy (eV/atom)
and atomic concentration at the roughening surface 
are shown
as a function of the $500$ eV Xe$^{+}$ ion impacts
and at $50^{\circ}$ incidence angle
for iob-beam assisted deposition (ibad) and without Fe impurities (noibad)
for the B20 phase of FeSi.
}
\label{fesi}
\end{center}
\end{figure}

\begin{figure}[hbtp]
\begin{center}
\includegraphics*[height=6cm,width=8cm,angle=0.]{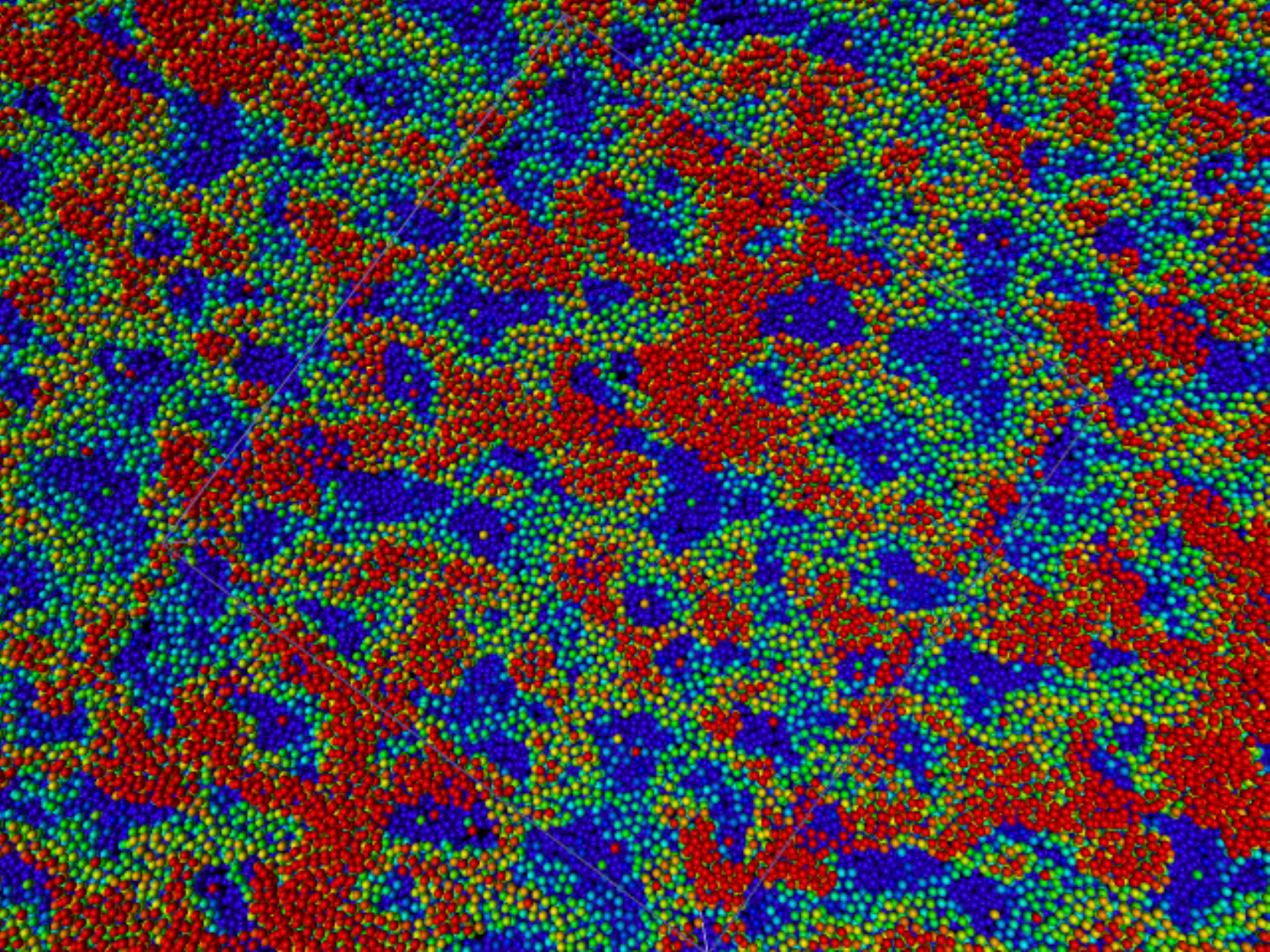}
\includegraphics*[height=6cm,width=8cm,angle=0.]{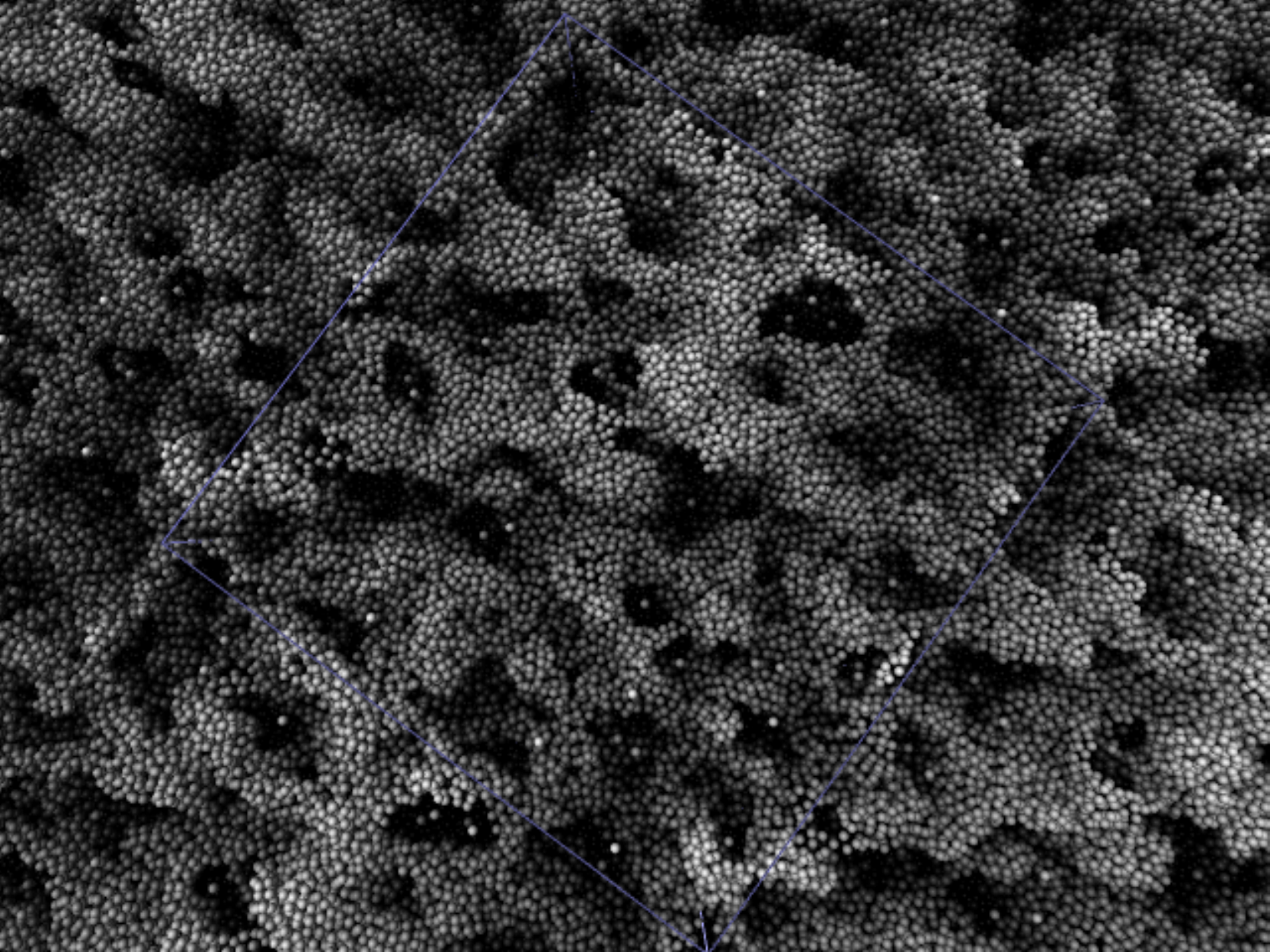}
\caption[]{
Color coded (rainbow and gray) height images of ion-sputtered surfaces of aSi
(noibad)
at $50^{\circ}$ $500$ eV $Xe^{+}$ impacts
The scanned area is $22.3 \times 22.3$ nm$^2$.
($2000$ repeated impacts randomly distributed over the whole area).
}
\label{xe50por}
\end{center}
\end{figure}

\begin{figure}[hbtp]
\begin{center}
\includegraphics*[height=6cm,width=8cm,angle=0.]{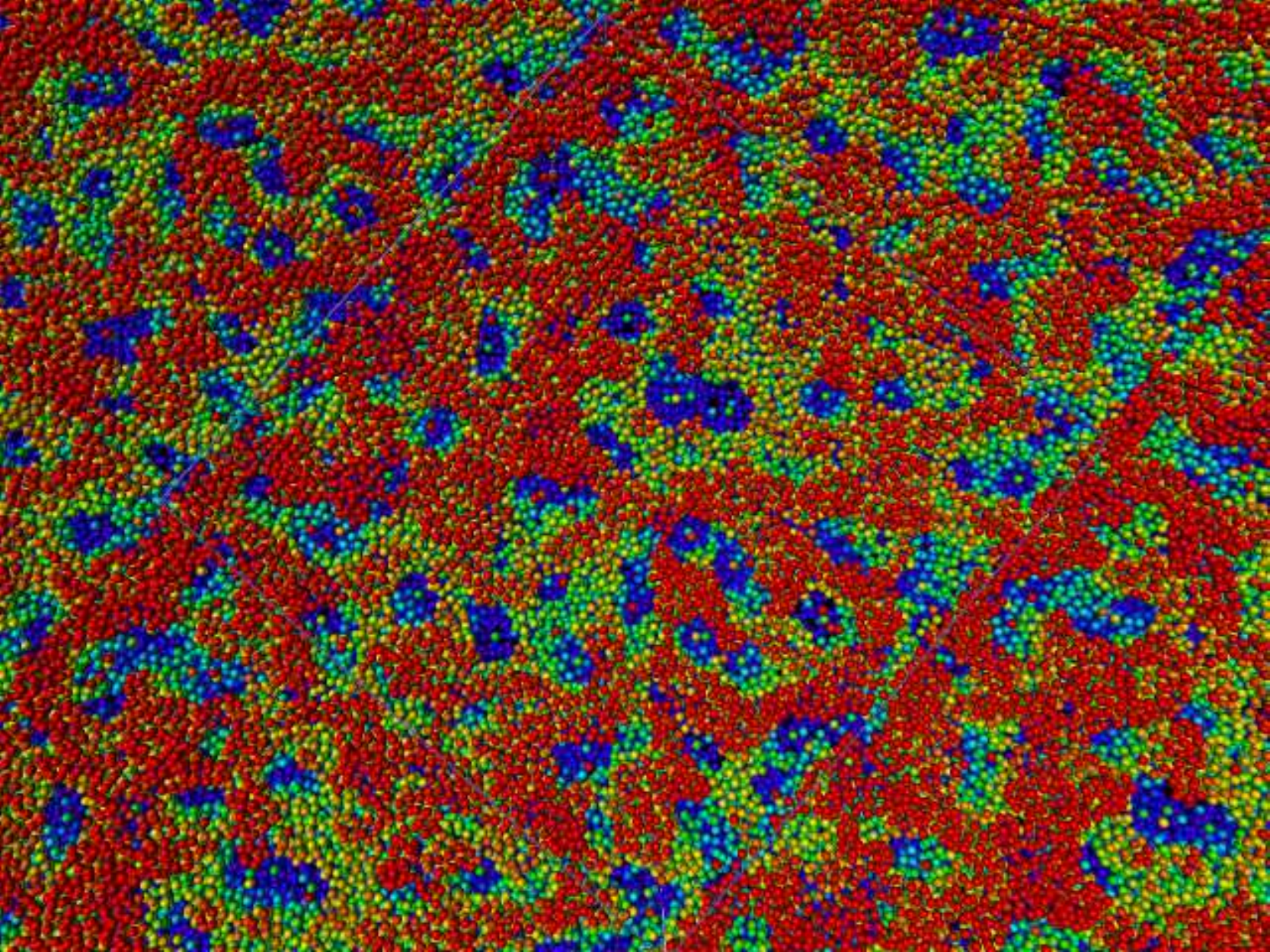}
\includegraphics*[height=6cm,width=8cm,angle=0.]{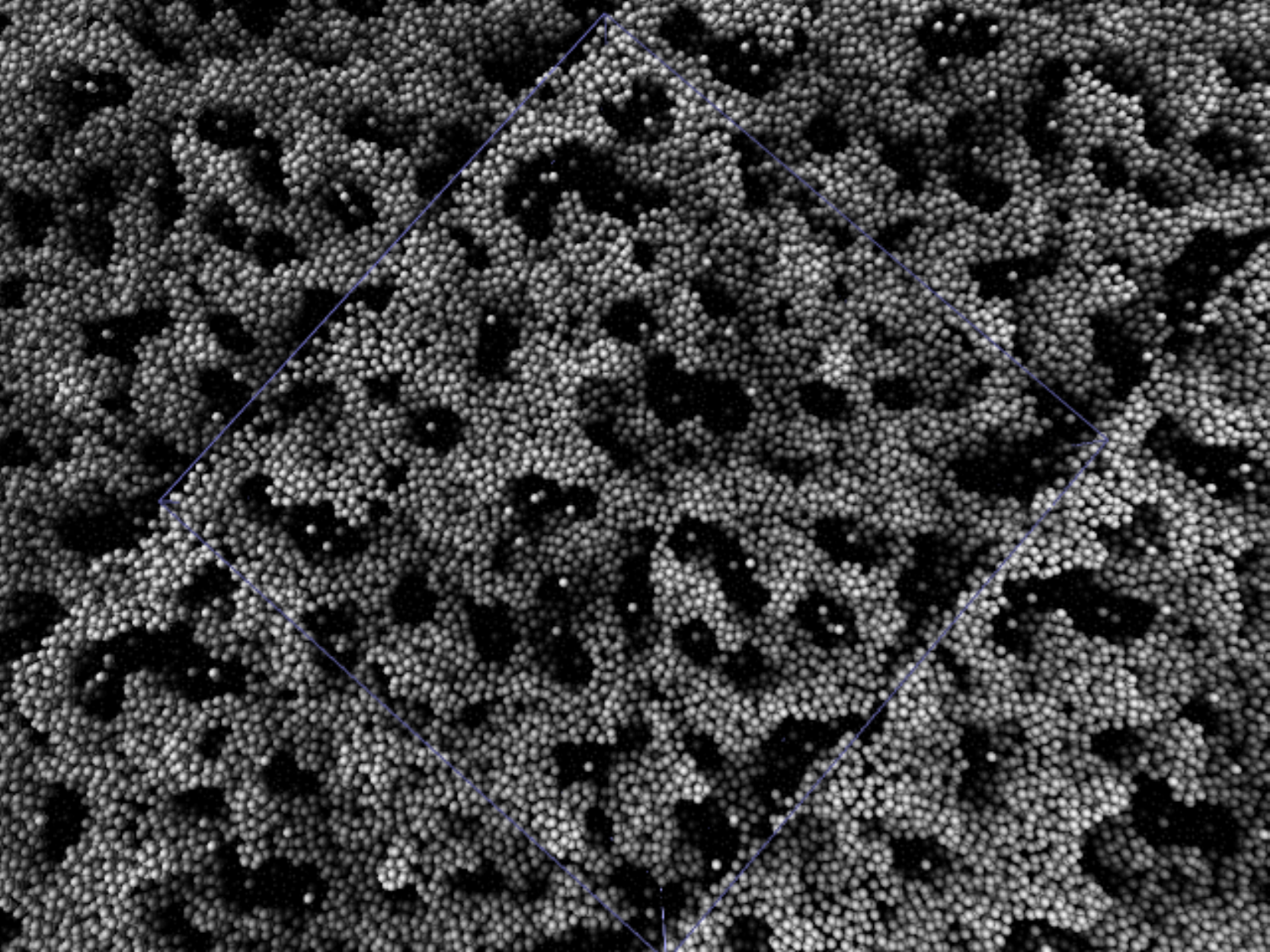}
\caption[]{
Color coded (rainbow and gray) height images of ion-sputtered surfaces of aSi
(noibad)
at $30^{\circ}$ $500$ eV $Xe^+$ impacts
The scanned area is $20 \times 20$ nm$^2$.
($2000$ repeated impacts randomly distributed over the whole area).
}
\label{xe30enl}
\end{center}
\end{figure}

  In order to avoid artificial effects in the topography, such
as overheating induced border walls of particles, the temperature
has been softly quenched to $300$ K after $2.5$ ps at each ion-impact
steps using temperature controll at the borders.
Although this way of temperature controll seems to be efficient, 
one has to keep in mind that radiation enhanced thermal diffusion has been excluded in our
computer experiments.
Ballistic dissusion and thermal spike (TS) is still accounted for in this modell
since collisional cascades and the subsequent heat spike terminates within
few ps \cite{Sule_SUCI05}.  Hence it is reasonable
to apply an artificial quenching process right after the TS to avoid the
overheating of the cell.
It is also known from simulations that
repeated ion-impacts induce damage accumulation which stimulates
partly radiation enhanced diffusion (RED) \cite{Sule_SUCI05}.
RED goes on a longer time scale than TS.
Its medium time scale range (from few ps up to a ns) could be followed
within our modell at 300 K.

 It has also been concluded in our recent paper \cite{Sule_JCP09}
that it is unlikely that thermal motion on a much longer time scale is responsible for
ion-induced nanopatterning and only minor smoothening effects on the
topography could be accounted for thermal diffusion.
Instead we emphasize that ballistic diffusion is responsible
for NP. We expect such ballistic motion of "hot" particles terminate within
$2.5$ ps.


This work is supported by the OTKA grant
 K-68312
from the Hungarian Academy of Sciences.
Support from the bilateral German-Hungarian
exchange program DAAD-M\"OB (Grant No. 37-3/2008)
and German Science Foundation (DFG research
group 845, project HE2137/4-1) is also
acknowledged. 
We wish to thank to K. Nordlund 
for helpful discussions and constant help.
The work has been performed partly under the project
HPC-EUROPA with the support of
the European Community using the supercomputing 
facility at CSC in Espoo.


\end{document}